\documentclass[twocolumn,floats,amsmath,amssymb,pre]{revtex4}

\usepackage{graphicx}

\begin{document}

\title{Structural characterization of ice polymorphs from self-avoiding 
       walks}
\author{Carlos P. Herrero}
\affiliation{Instituto de Ciencia de Materiales de Madrid,
         Consejo Superior de Investigaciones Cient\'{\i}ficas (CSIC),
         Campus de Cantoblanco, 28049 Madrid, Spain }
\date{\today}

\begin{abstract}
Topological properties of crystalline ice structures are studied 
by means of self-avoiding walks on their H-bond networks.
 The number of self-avoiding walks, $C_n$, for eight ice polymorphs 
has been obtained by direct enumeration up to walk length $n = 27$. 
This has allowed us to
determine the `connective constant' or effective coordination number 
$\mu$ of these structures as the limit of the ratio $C_n / C_{n-1}$ 
for large $n$.  This structure-dependent parameter $\mu$ is related 
with other topological characteristics of ice polymorphs, such as the 
mean and minimum ring size, or the topological density of 
network sites. 
A correlation between the connective constant and the configurational
entropy of hydrogen-disordered ice structures is discussed. 
\end{abstract}

\maketitle

\section{Introduction}

Water exhibits a wide variety of solid phases, which are referred to as
forms of `ice'. Most of these phases are produced by the application of
high pressures, which yields a denser packing of water molecules than in
the usual hexagonal ice Ih.
Thus, sixteen different crystalline ice phases have been
identified so far \cite{pe99,du10,ba12},
and their stability range in the temperature-pressure phase 
diagram has been investigated for several decades.
Some of their properties lack, however, a complete understanding,
mainly due to their peculiar structure, where 
hydrogen bonds between adjacent molecules give rise to
the so-called `water anomalies' \cite{ei69,pe99,ro96}.

In the known ice phases except ice X, water keeps its molecular character, 
building up a network connected by H-bonds. 
In this network each water molecule is surrounded by four others,
in such a way that its orientation with respect to the neighboring ones 
fulfills the so-called Bernal-Fowler ice rules \cite{be33,pe99}.
These rules allow for the presence of orientational disorder in the 
water molecules, which causes that  in several ice phases 
hydrogen atoms present a disordered spatial distribution, as 
indicated by a fractional occupancy of their lattice sites. 
Thus, hexagonal ice Ih, the stable phase of solid water under normal 
conditions, displays hydrogen disorder compatible with the ice rules, 
whereas other phases such as ice II are H-ordered \cite{sa11,si12}. 

Given the number of ice structures, a unifying classification
can help to deeper understanding of their specific 
properties \cite{ma09,sa11}.
For crystalline solids, classification schemes usually rest on the 
space symmetry, short-range atomic environments, or geometrical aspects 
of packing of structural units.
These classification procedures have a geometrical nature, as their
main criteria are geometrical characteristics of crystal
structures \cite{pe72,we86,li85,bl00}.
These geometrical classification methods, however, can be hardly
applicable to find relations between solids whose structures
are somewhat distorted.
A possible alternative consists in using classification schemes relying on
topological criteria. This means centering attention mainly on the 
organization of interatomic bonds in a crystal structure as a basic 
criterion for a crystal-chemical analysis.
In this line, topological properties of crystalline solids have been
taken into account along the years to describe properties of different 
types of materials \cite{pe72,li85,we86}.
For the ice polymorphs, a discussion of different network topologies 
and the relation of ring sizes in the various phases with the crystal
volume was presented by Salzmann {\em et al.} \cite{sa11}.
Topological studies of three-dimensional (3D) hydrogen-bonded frameworks 
in organic crystals have also helped to classify this kind of
structures \cite{ba07}.
Moreover, graph theory has been used to study isotypism and
order/disorder problems in crystal structures \cite{bl00,si12,kn08}.

A direct and rather simple topological classification of ice structures 
can be based on structural rings, i.e., loops of water molecules 
characteristic of each polymorph \cite{sa11}.
A more elaborate procedure can be based on the so-called `coordination
sequence', defined as a series of numbers $\{N_k\}$ ($k$ = 1, 2, ...), 
where $N_k$ is the number of sites located at a topological distance $k$ 
from a reference site \cite{co97,gr96,he94,eo02}.
The coordination sequence can be used to define a topological
density, as a structural characteristic related to the
increase in the number $N_k$ of sites accessible through $k$ links in
a given structure \cite{br93,gr96,eo04}.
Note that these concepts are based only on the topology of the considered 
network, and are not affected by lattice distortions or other structural 
factors.
These concepts have been recently applied to ice polymorphs, allowing
us to find a correlation between topological density and
volume \cite{he13b}.

  In this paper, a different way to characterize the ice polymorphs
is introduced. Namely, it is based on self-avoiding walks 
in the corresponding structures. 
A self-avoiding walk (SAW) is a sequence of moves on a network that does 
not visit any node more than once.
Contrary to unrestricted walks, SAWs contain implicit information on
the topology of the considered network, as they are sensitive to
characteristics such as the number and size of loops present in the
structure.
A particularly interesting parameter is the so-called connective constant 
or effective coordination number of the ice networks, which can be
calculated from the long-distance behavior of the number of possible 
SAWs in the corresponding structures. These concepts are explained in 
detail in the following section.

The purpose of this paper is twofold. On one side, knowledge of the
connective constants of ice structures is interesting from a basic
point of view for their comparison with other crystal structures,
for which this kind of statistical-mechanics questions have been
analyzed in detail.
On the other side, they are relevant for a topological and physico-chemical
characterization of ice phases, allowing us to connect structural and
thermodynamic properties of this type of solids.
In particular, the configurational entropy associated to the hydrogen 
distribution on the available lattice sites is known to depend on the
ice structure \cite{he13,ma04b,be07}, so that a search for structural
variables suitable to quantify in some way such a dependence is 
worthwhile in the context of ice thermodynamics.

\section{Computational method}

In order to study SAWs for the different ice polymorphs, we consider 
each structure as defined by the positions of the oxygen atoms.
One has thus a network, where the nodes are O sites, and the 
links are H-bonds between nearest neighbors.
The network coordination is four ($z = 4$), which gives a total of 
$2N$ links, $N$ being the number of nodes. 
We implicitly assume that on each link
there is one H atom, but its consideration is not relevant for our
present purposes.

\begin{figure}
\vspace{-1.4cm}
~\hspace{-1.9cm}
\includegraphics[width=11cm]{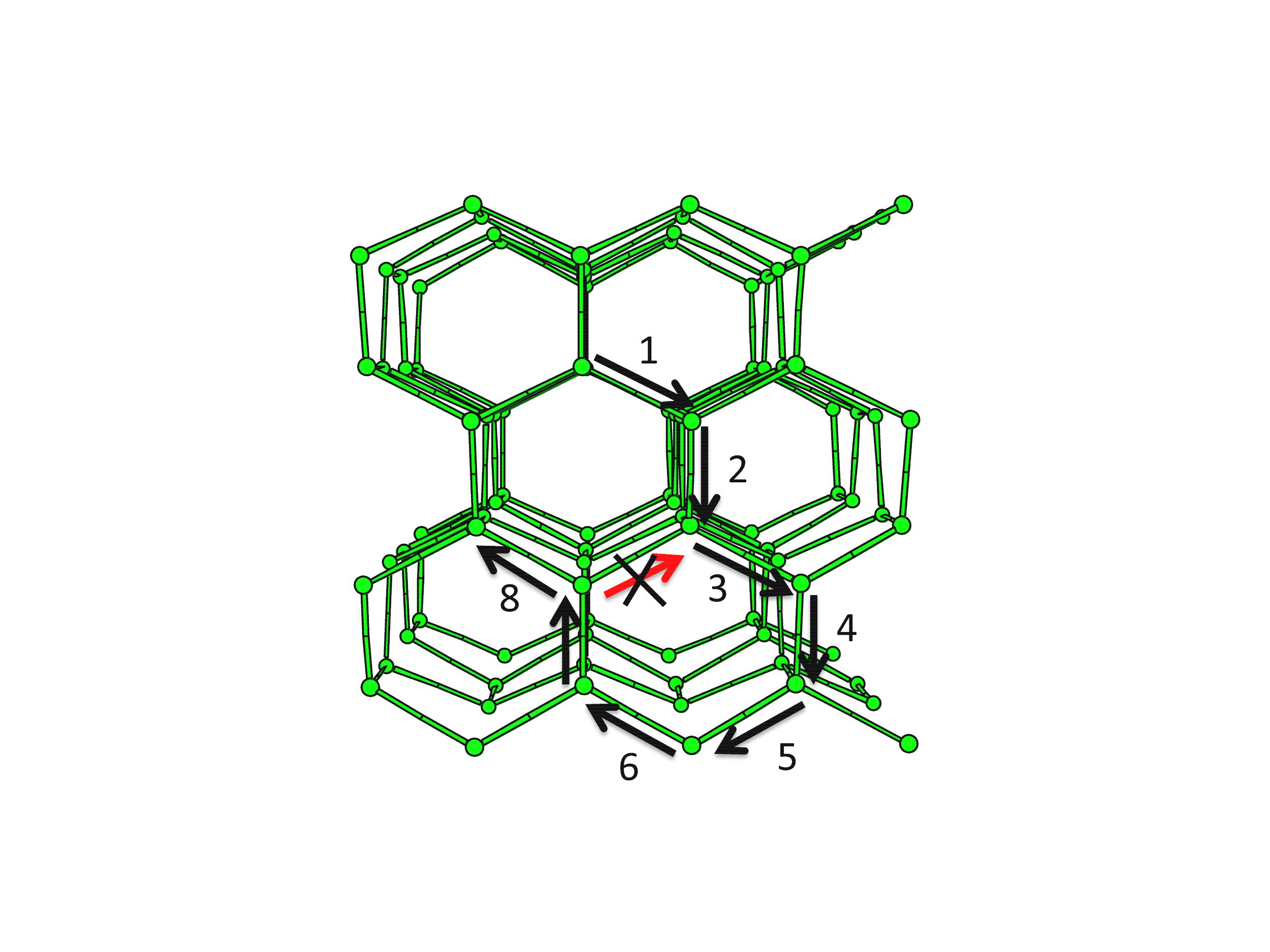}
\caption{
Sketch of the ice Ih structure with an eight-step self-avoiding walk.
A cross indicates a non-allowed step.
}
\label{f1}
\end{figure}

 A self-avoiding walk on an ice network is defined as a walk
in the simplified structure which can never intersect itself. 
On a given network, the walk is restricted to move to
a nearest neighbor site during each step, and the self-avoiding
condition constrains the walk to occupy only sites which have not been
previously visited in the same walk \cite{do69,bi88,je04}.
We illustrate the application of this definition to ice Ih in Fig.~1,
where an eight-step SAW is shown. Note that the link indicated with a
cross is not available for step $n$ = 8, as the walk would reach a node
already visited in an earlier step.

SAWs have been used in condensed-matter science for several purposes.
For instance, they were employed for modeling the large-scale properties 
of long-flexible macromolecules in solution and adsorbed on 
surfaces \cite{bi88,ge79,ca06,ry11}, as well as
for the study of polymers trapped in confined regions, gel
electrophoresis, and size exclusion chromatography, which deal with the
transport of polymers through membranes with very small 
pores \cite{le89,ja00,va98}. They have been also employed in
the analysis of critical phenomena in lattice
models \cite{do69,kr82,kr88,ma00,ca05}, and
to study complex networks \cite{he95b,he03c,he05b}.
Moreover, SAWs with multiple site weightings and restrictions
have been discussed in the literature \cite{pr01,kr06}.

Given a network and a site $i$ in it, we will call $C_n^i$ the number 
of different SAWs of length $n$ starting from this site.
For networks where all sites are topologically equivalent, the sequence
$\{ C_n^i \}_{n=1}^{\infty}$ will coincide for all nodes, but 
in general, sequences corresponding to different nodes in a network 
may be different. (Note that for all sequences mentioned in the following 
$n$ is understood to run from 1 to $\infty$, although not explicitly
indicated.)  
It is important to recall that crystallographically equivalent sites 
are always topologically equivalent, but sites non-equivalent 
crystallographically may be topologically equivalent or not \cite{he13b}.
Then, we define an average sequence
$\{ C_n \}$ for each network, where for a given $n$, $C_n$ is obtained
by averaging the $C_n^i$ values for the oxygen sites in the unit cell:
\begin{equation}
   C_n = \frac1M \,  \sum_i  m_i \, C_n^i  \; .
\label{cn2}
\end{equation}
Here $m_i$ is the multiplicity of site $i$ in the unit cell and
$M = \sum_i m_i$.
For ice structures including oxygen sites topologically non-equivalent
(e.g., ices III, IV, V, VI, and XII; see Ref.~\cite{he13b}),
relative differences between $C_n^i$ values corresponding to different
sites in a given structure decrease fast with the walk length $n$. In
fact, the relative difference is about 1\% for $n$ = 25, and becomes
negligible in the large-$n$ limit.

It is remarkable that universal constants are known to describe
some properties of self-avoiding walks. These constants depend on the
network dimension, and have been discussed in detail in the
literature \cite{pr91,mc76}.
Other parameters controlling the long-distance behavior of SAWs are
network-dependent, and can be used to characterize different networks 
with the same dimension.
Analytical expressions describing the asymptotic behavior of SAWs for
large $n$ are significantly different from those of unrestricted walks.
 It is particularly interesting the dependence of $C_n$ on the number 
of steps $n$ for long walks.
Its asymptotic behavior for large $n$ is known to be  
given by \cite{pr91,mc76,ra85}
\begin{equation}
         C_n \sim   n^{\gamma - 1}   \mu^n  \hspace{2mm} ,
\label{cn}
\end{equation}
where $\gamma$ is a critical exponent which takes a value 
$\approx 7/6$ for 3D structures \cite{mc76,ca98,ch02}, and
$\mu$ is the so-called `connective constant' or effective coordination
number of the corresponding structure \cite{bi88,mc76,ra85}.

Defining the ratio
\begin{equation}
    x_n = \frac{C_n}{C_{n-1}}
\label{xn}
\end{equation}
one has
\begin{equation}
    x_n \approx \mu \, \left( 1 + \frac{1}{n-1} \right)^{\gamma - 1} 
         \xrightarrow{\scriptscriptstyle n\to\infty} \mu
\label{xn2}
\end{equation}
so that the (non-universal) parameter $\mu$ can be obtained as the
large-$n$ limit of the sequence $\{ x_n \}$.
The connective constant depends upon the particular topology of 
each structure, and has been determined very accurately for standard 
3D lattices.
 In particular, for the diamond structure (with the same
topology as cubic ice Ic), one has $\mu$ = 2.8790 \cite{ch02,gu89b}.
In general, for structures with coordination number $z$, one has
$\mu \le z - 1$ \cite{bi88}.

For some ice networks, the convergence of $x_n$ to the connective
constant $\mu$ is rather smooth, but in other cases an odd-even
alternation is present. For this reason, it is convenient 
to average out this effect partially by evaluating the average values
\begin{equation}
   y_n = \frac12 \,  (x_n + x_{n+1})
\label{yn}
\end{equation}
and then studying the convergence of the sequence $\{ y_n \}$ for
$n \to \infty$.  This is the procedure that will be employed here.
We have verified that the connective constant $\mu$ so obtained from the
average sequence $\{ C_n \}$ coincides within error bars with that 
yielded by averaging the values derived independently from the
sequences $\{ C_n^i \}$ of different network nodes.

This procedure gives a good estimate of the connective constant, but
other, more elaborate, numerical methods such as differential approximants 
are known to yield more accurate estimates for the extrapolated values of
$\mu$ from the sequence $\{ C_n \}$ \cite{gu89,je04}. Such a high
precision, required for some studies in statistical physics, is not 
necessary for our present purposes. In fact, the estimated error bar 
of the $\mu$ values derived from our extrapolation of the sequence
$\{ y_n \}$ is $\pm 2 \times 10^{-4}$. A check of this extrapolation
method is provided by the ice Ic network, that is equivalent in our
context to the diamond structure. The parameter $\mu$ derived here for
ice Ic is consistent with the value found for diamond in earlier 
estimations \cite{ch02,gu89b} (see below).

The sequence $\{ C_n \}$ for a given ice structure depends only on 
the topology of the network, and not on the 
actual symmetry, cell parameters, or other structural data. 
In particular, it is not affected by the ordering of H atoms,
and structures such as those of ices Ih and XI (the former
being H-disordered and the latter H-ordered), with the same topology,
will have the same sequence $\{C_n\}$.
At present, there are five other known pairs of ice structures sharing 
the same topology,
and related one to the other through an order/disorder phase
transition.  Then, one has six pairs of ice polymorphs:
Ih-XI, III-IX, V-XIII, VI-XV, VII-VIII, and XII-XIV \cite{sa11},
where the first polymorph in each pair is H-disordered and
the second is H-ordered.
For our present purposes, we will only refer in the following to the
disordered case, but understanding that it represents both
members of the corresponding pair of ice polymorphs. In addition to these, 
there exist other ice structures for which no pair has been found:
ice Ic (H-disordered), ice II (H-ordered), and ice IV (H-disordered).
Finally, ice X is topologically equivalent
to ices VII and VIII, but with the important difference that in 
ice X hydrogen atoms lie midway between oxygen atoms, so
that water molecules lose in fact their own entity (this is however
unimportant for our present considerations).
Thus, one has 9 topologically different ice structures.
We note that the network associated to ice VII
(as well as ices VIII and X) is composed of two
interpenetrating but disjoint subnetworks, each of them
equivalent to the ice Ic network. 
There is another case, the pair VI-XV, for which the network also 
consists of two disjoint subnetworks, but they are not equivalent to 
the network of any other known ice polymorph.

\begin{table}[ht]
\caption{Crystal system and space group for the ice polymorphs
considered in this work, along with the references from where
crystallographic data were taken.  \\}
\centering
\setlength{\tabcolsep}{10pt}
\vspace*{0.0cm}
\begin{tabular}{cccc}
  \hline \hline \vspace*{-0.2cm} \\
  \vspace*{0.1cm}
   Phase & Crystal system & Space group &  Ref. \\
   \hline \vspace*{-0.2cm} \\
   Ih    & Hexagonal    & $P6_3$/$mmc$, 194 &  \cite{pe57} \\
   Ic    &   Cubic      & $Fd\bar{3}m$, 227 &  \cite{ko44} \\
   II    & Rhombohedral & $R\bar{3}$, 148   &  \cite{ka71} \\
   III   & Tetragonal   & $P4_12_12$, 92    &  \cite{lo00} \\
   IV    & Rhombohedral & $R\bar{3}c$, 167  &  \cite{en81} \\
   V     & Monoclinic   & $A2/a$, 15        &  \cite{lo00} \\
   VI    & Tetragonal   & $P4_2/nmc$, 137   &  \cite{ku84} \\
   VII   & Cubic        & $Pn\bar{3}m$, 224 &  \cite{ku84} \\
   XII   & Tetragonal   & $I\bar{4}2d$, 122 &  \cite{lo98} \\
   &&& \vspace*{-0.3cm} \\
   \hline \hline
\end{tabular}
\label{tb:ice_crystall_systems}
\end{table}

For the above reasons we will study here the ice polymorphs given 
in Table~I, where we indicate the crystal system, space group, and
the reference from where we took the crystal data to generate the 
corresponding supercells.
In general, a study of the asymptotic behavior of the sequence
$\{ C_n \}$ for a given network requires the generation of large
supercells.  For the ice structures considered in this work, supercells
including around $10^5$ oxygen sites were generated.
We have calculated the numbers $\{ C_n \}$ of SAWs in such supercells 
by exact enumeration of the possible walks up to $n = 27$, using
a backtracking algorithm \cite{ra87}. 

\begin{figure}
\vspace{-1.0cm}
\includegraphics[width= 8cm]{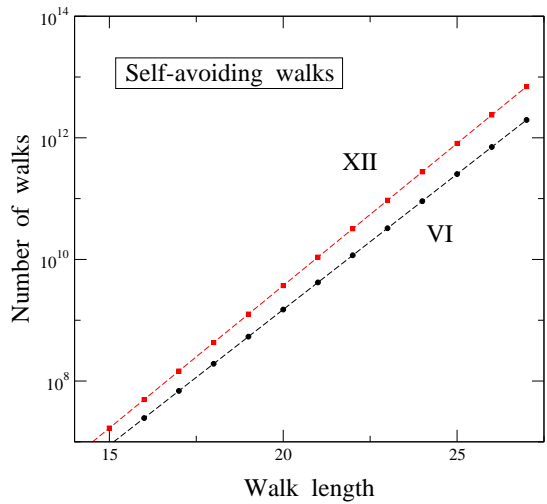}
\vspace{-0.3cm}
\caption{
Linear-log plot of the number of self-avoiding walks $C_n$ on the
structures of ice VI (circles) and XII (squares).
}
\label{f2}
\end{figure}

\section{Results and discussion}

 The mean numbers $C_n$ of SAWs corresponding to different ice
structures are presented in Table~II for three walk lengths: 
$n$ = 5, 10, and 15.  
Note that in some cases $C_n$ is not an integer number, since 
$C_n^i$ corresponding to different starting sites $i$ are different
[see Eq.~(\ref{cn2})].
We have checked that the sequence $\{ C_n \}$ obtained for ice Ic
coincides with that presented earlier for diamond up to $n = 27$
\cite{ch02,gu89b}.
The number of walks
$C_n$ increases fast for rising $n$, as expected from the exponential
part $\mu^n$ in Eq.~(\ref{cn}) for $\mu > 1$.
The increase in $C_n$ as a function of the walk length $n$ is shown 
in Fig.~2 for ices VI and XII, in a linear-logarithmic plot.
Among the considered ice polymorphs, these two present the smallest and
largest $C_n$ values, respectively, and for other crystalline ice
structures, $C_n$ values lie between those of ices VI and XII.
Comparing results for these two structures, we find a ratio
$C_{27}(\rm XII) / C_{27}(\rm VI)$ = 3.52 for SAWs of length 
$n = 27$, the largest walks considered here.
This ratio between $C_n$ values of different structures increases
as $n$ is raised. In fact, 
according to Eq.~(\ref{cn}), $C_n(\rm XII) / C_n(\rm VI)$
should increase for large $n$ as $(\mu_{\rm XII} / \mu_{\rm VI})^n$.
The dependence of $\log_{10}C_n$ on $n$ shown in Fig.~2 is roughly
linear, showing the leading contribution of the exponential term 
$\mu^n$ in Eq.~(\ref{cn}).
In general, for the structures under consideration, one expects
\begin{equation}
   \log_{10} C_n - \log_{10} C_{n-1} = \log_{10} x_n   
     \xrightarrow{\scriptscriptstyle n\to\infty}   \log_{10} \mu
\end{equation}
Thus, the slope of $\log_{10} C_n$ vs $n$ shown in Fig.~2 converges 
for large $n$ to $\log_{10} \mu$.

\begin{table*}[ht]
\caption{Average number $C_n$ of SAWs for walk length $n$ =  5, 10, and
15, for different ice structures, along with the corresponding connective
constant $\mu$. Error bars of $\mu$ values are $\pm 2 \times 10^{-4}$.
$L_{min}$ and $\langle L \rangle$ are the minimum and mean ring size
for each ice structure.
$a$ is the coefficient of the quadratic term in the coordination
sequence, $N_k \sim a k^2$, as in Eq.~(\ref{nk}).
Data for ice VII are not given, since they coincide with those of
ice Ic.  \\ }
\centering
\setlength{\tabcolsep}{10pt}
\vspace*{0.5cm}

\begin{tabular}{ c c c c c c c c}
  \hline \hline \vspace*{-0.2cm} \\
  \vspace*{0.1cm}
 Network &  $C_5$  &  $C_{10}$ &  $C_{15}$ & $\mu$  & $a$ &
        $\langle L \rangle$  &  $L_{min}$  \\
   \hline \vspace*{-0.2cm} \\
 Ih        &  324 &  70188 & 14776128    &  2.8793 &  2.62 & 6    & 6 \\
 Ic        &  324 &  70188 & 14774652   &   2.8792 &  2.50 & 6    & 6 \\
 II        &  324 &  73032 & 15996858   &   2.9049 &  3.50 & 8.52 & 6 \\
 III       &  317.3 & 68346.7 & 14156657.3 & 2.8714 & 3.24 & 6.67 & 5 \\
 IV        &  324   &  74241  & 16537432.5 & 2.9162 & 4.12 & 9.04 & 6 \\
 V         &  308.3 & 63877.4 & 12886447.7 & 2.8596 & 3.86 & 8.36 & 4 \\
 VI        &  284   & 51021.6 & 8855288.8  & 2.7706 & 2.00 & 6.57 & 4 \\
 XII       &  324   & 75245.3 &  16832260  & 2.9179 & 4.26 & 7.6  & 7 \\
   &&& \vspace*{-0.3cm} \\
\hline  \vspace*{-0.2cm}  \\
 Square    &  284   & 44100   &  6416596   & 2.6382 & 0.0  & 4    & 4 \\
 Bethe     &  324   & 78732   &  19131876  & 3.0000 &  --  & --   & -- \\
   &&& \vspace*{-0.1cm} \\
   \hline \hline
\end {tabular}
\label{tb:coord_seq}
\end{table*}

To obtain insight into the differences between $C_n$ values for the
different ice polymorphs, it is convenient to consider the topology of
the corresponding crystal structures.
Thus, given the variety of ice structures, it is not strange that 
some topological characteristics display clear differences
between different polymorphs, as could be
expected from the pressure/temperature range where they are stable.
In this line, a first characteristic observable in the ice networks
is the presence of loops (rings of water molecules),
which define the particular topology of each structure.
The statistics of rings in ice polymorphs has been considered in detail
elsewhere \cite{sa11,he13b}.
In particular, the mean ring size, $\langle L \rangle$,
can be useful to connect topological and thermodynamic properties of
these ice structures.
In Table~II we present $\langle L \rangle$ along with the minimum ring 
size $L_{min}$ for the various polymorphs.

For a loop-free network (Bethe lattice or Cayley tree \cite{zi79,st90}),
the number of SAWs is given by
\begin{equation}
  C_n^B = z \, (z-1)^{n-1}
\label{cnb}
\end{equation}
where $z$ is the number of nearest neighbors (degree or connectivity
in the language of graph theory), assumed to be the same for all sites.
For actual crystal structures, with networks including loops, $C_n$ will
be in general smaller than $C_n^B$, and for a given $n$ the
number of $n$-step SAWs is affected by the presence of rings of size  
$L \leq n$.  For example,
for our networks with $z = 4$ the maximum number $C_5$ of 5-step SAWs 
is that corresponding to the Bethe lattice: $C_5^B$ = 324.
Then, the lower values of $C_5$ for some ice networks (see Table~II) 
are due to the presence of rings including five or less nodes. 
This happens for ices V and VI which contain four-membered rings
($L_{min} = 4$) and ice III with $L_{min} = 5$.
For longer SAWs (larger $n$), loops with higher number
of nodes also contribute to reduce the number $C_n$ with respect to
that of the Bethe lattice. 

\begin{figure}
\vspace{-1.0cm}
\includegraphics[width= 8cm]{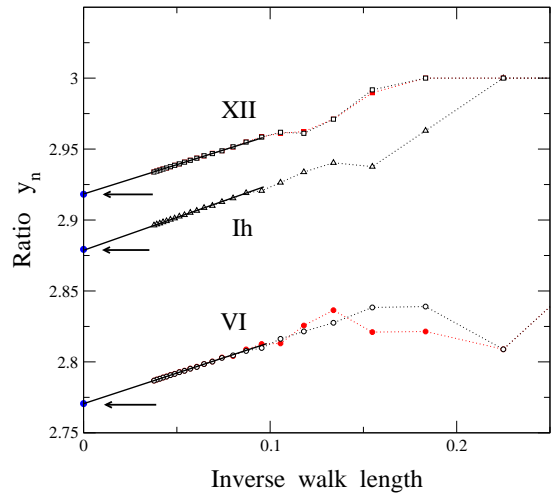}
\vspace{-0.3cm}
\caption{
Ratio $y_n$ vs inverse walk length for
ice VI (circles), Ih (triangles), and XII (squares).
For ice VI and XII, results for two topologically different starting
sites are shown (open and solid symbols), as derived from the
corresponding sequence $\{ C_n^i \}$.
For each structure an arrow indicates the value of the connective
constant $\mu$ obtained from an extrapolation of $y_n$
for $1/n \to 0$.
}
\label{f3}
\end{figure}

The connective constant $\mu$ has been obtained for the considered ice
structures from the corresponding sequences $\{ C_n \}$ by using 
the procedure described in Sect.~II. 
In Fig.~3 we present the mean values $y_n$ defined in Eq.~(\ref{yn}) 
for three different
structures (ices Ih, VI, and XII) as a function of the inverse walk
length $1/n$. In fact, for ices VI and XII we show separately the 
values derived
from the ratios $C_{n+1}^i/C_n^i$ corresponding in each case to two
topologically non-equivalent sites (open and solid symbols). 
Although results for those non-equivalent sites can be clearly different
for small $n$ values, for larger $n$ they converge for each ice
polymorph to the same limit. This should be expected, as for large
length $n$, the walk looses memory of the starting node. 
The resulting values for the connective constant $\mu$ of the
considered structures are given in Table~II.
We estimate for these $\mu$ values an error bar of 
$\pm 2 \times 10^{-4}$, due basically to the extrapolation  
$n \rightarrow \infty$. 
A check of our procedure is provided by the diamond structure, 
topologically equivalent to the ice Ic network, and for which the
connective constant is known to be $\mu$ = 2.8790 \cite{mc76}.
This value coincides within error bars with our result for ice Ic:
$\mu$ = 2.8792(2).
For comparison with results for the ice networks, note that for the 
loop-free Bethe lattice with $z$ = 4, one has $\mu$ = 3, 
as can be straightforwardly derived from Eq.~(\ref{cnb}).
We also include in Table~II the connective constant for the
two-dimensional (2D) square lattice: $\mu$ = 2.6382 \cite{gu01}
(also with $z = 4$).

\begin{figure}
\vspace{-0.6cm}
\includegraphics[width= 8cm]{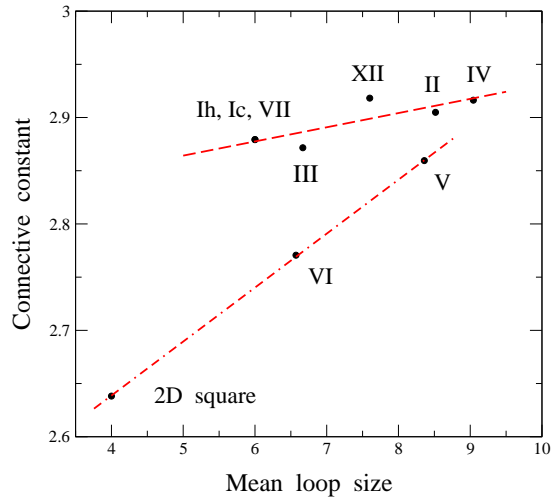}
\vspace{-0.3cm}
\caption{
Connective constant $\mu$ vs mean ring size $\langle L \rangle$ for
crystalline ice structures.
Labels indicate the ice structure corresponding to each data
point.  Ices Ih, Ic, and VII appear as a single point.
The dashed-dotted line is a fit for the structures including
four-membered rings (2D square lattice, ices V and VI).
The dashed line is a least-square fit for the other ice structures.
}
\label{f4}
\end{figure}
 
As indicated above, the connective constant $\mu$ is a purely
topological concept, and therefore should in some way be related with
other (more basic) topological aspects of the ice structures.
The first quantity to be compared with $\mu$ will be the mean loop size 
$\langle L \rangle$.
In Fig.~4 we present the connective constant of various ice polymorphs
vs $\langle L \rangle$. For the sake of comparison, we include a data
point for the 2D square lattice, where all loops have 
$L = 4$.
One first observes in this plot a tendency of the connective constant 
$\mu$ to increase for rising $\langle L \rangle$.
There appears, however, an appreciable dispersion of the data points
for different ice polymorphs, indicating a low correlation between both
quantities.
A more careful observation of the data shown in Fig.~4 reveals that
the points corresponding to the ice structures are roughly aligned,
with the exception of ices V and VI that appreciably depart from the
general trend. Common to these two ice polymorphs is that they are the
only ones including four-membered rings ($L_{min} = 4$).
In fact, one observes a good linear correlation between $\mu$ and 
the mean ring size $\langle L \rangle$ for these structures and the 2D
square lattice (dashed-dotted line).
The dashed line in Fig.~4 is a linear fit to the data points of ice
structures not including four-membered rings. For these structures
we have $L_{min} > 4$, i.e., $L_{min} = 5$ for ice III, 
$L_{min} = 6$ for ices Ih, Ic, II, and IV, and
$L_{min} = 7$ for ice XII.
We observe in fact that the point corresponding to ice III lies below
the dashed line, whereas that corresponding to ice XII appears above
it. This is in line with the trend found for ices V and VI, since 
$L_{min}({\rm III}) < 6 < L_{min}({\rm XII})$.

From these results
we conclude that the connective constant tends to increase as the ring
mean size increases, but the presence of four-membered rings strongly
affects the actual value of $\mu$. Thus, the connective constant for
ice VI is clearly lower than that of ice III, in spite of the fact that
both polymorphs have very similar values for $\langle L \rangle$.
This influence of four-membered rings decreases, however, for larger 
$\langle L \rangle$, as can be seen by comparing $\mu$ values for ices 
II an V.

As mentioned in the Introduction,
the network topology can be characterized by the so-called
coordination sequences $\{N_k\}$ ($k$ = 1, 2, ...), where $N_k$
is the number of sites at a topological distance $k$ from a reference 
site \cite{he13b}. For 3D structures, $N_k$ increases at
large distances as:
\begin{equation}
         N_k \sim a \, k^2     \hspace{3mm}  ,
\label{nk}
\end{equation}
where $a$ is a network-dependent parameter. $N_k$ increases
quadratically with $k$ just as the surface of a sphere increases
quadratically with its radius.
For structures including topologically non-equivalent sites, the actual
coordination sequences corresponding to different sites may be
different, but the coefficient $a$ coincides for all sites in a given
structure \cite{he13b}.
The parameter $a$ can be used to define a `topological density' for
ice polymorphs, $\rho = w \, a$, where $w$ is the number of 
disconnected subnetworks in the considered network \cite{he13b}.
Usually $w = 1$, but for ice structures including two interpenetrating 
networks (as ices VI and VII) one has $w = 2$.

\begin{figure}
\vspace{-1.0cm}
\includegraphics[width= 8cm]{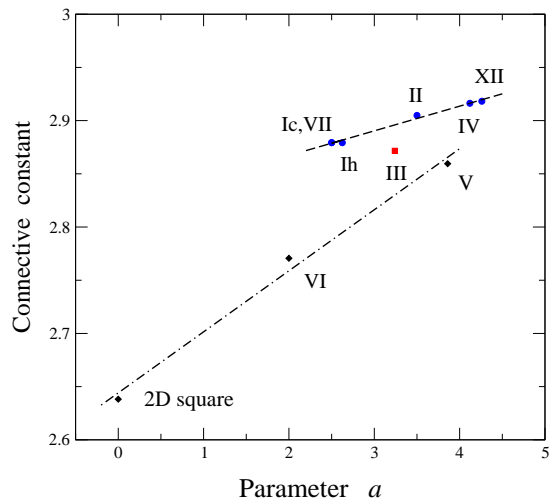}
\vspace{-0.3cm}
\caption{
Connective constant $\mu$ vs parameter $a$ of coordination sequences
for different ice polymorphs. Labels indicate the ice structure
corresponding to each data point.
Ices Ic and VII are represented by a single point, since they have the
same coordination sequence (parameter $a$) and connective constant.
The dashed-dotted line is a fit for the structures
including four-membered rings (2D square lattice, ices V and VI).
The dashed line is a least-square fit for ice polymorphs with minimum
ring size $L_{min} > 5$.
}
\label{f5}
\end{figure}

In Fig.~5 we display the connective constant $\mu$ of ice polymorphs
vs the coefficient $a$ derived from the coordination sequences.
As in Fig.~4, we also include a point for the 2D square lattice.
We find again that $\mu$ increases for increasing parameter $a$, but
there appears a dispersion in the data points, which can be associated
to the minimum ring size of the considered ice structures.
In particular, ice polymorphs including four-membered rings
($L_{min} = 4$) depart appreciably from the general trend of the other
structures, similarly to the correlation between $\mu$ and the mean
loop size $\langle L \rangle$ shown in Fig.~4.
Then, we also present in Fig.~5 two different fits. On one side, 
the dashed-dotted line is a linear fit for the structures with
$L_{min} = 4$, including the 2D square lattice. On the other side,
the dashed line is a least-square fit for the ice polymorphs with
$L_{min} > 5$, where all of them have $L_{min} = 6$ except ice XII,
for which $L_{min} = 7$ (see Table~II). 
For the remaining ice polymorph (ice III), with $L_{min} = 5$, the
corresponding point (full square) lies on the region between both
lines in Fig.~5.

One of the goals of the present work consists in finding relations
between topological characteristics and thermodynamic properties of ice
polymorphs.
In this line, it has recently been shown that the network topology
plays a role to quantitatively describe the configurational entropy
associated to hydrogen disorder in ice structures \cite{he13}.
This has been in fact known after the work by Nagle \cite{na66},
who calculated the entropy for ices Ih and Ic. This author showed 
that the actual structure is relevant for this purpose, by comparing 
his results with Pauling's value for the configurational entropy,
derived in a simple approximation for a loop-free network (Bethe
lattice).
It has been recently shown that a simple procedure based on
thermodynamic integration yields reliable results for the
configurational entropy, $s$, of other ice structures with
hydrogen disorder \cite{he13}.
This method was applied to ice VI, for which an entropy per site
$s$ = 0.4214(1) was found, vs $s$ = 0.4107(1) for ice Ih and
Pauling's value $s$ = 0.4055 \cite{pa35}. Note that $s$ is 
dimensionless and can be converted to the physical configurational 
entropy $S$ as $S = N k_B s$, with $N$ the number of water molecules 
(nodes in the simplified network) and $k_B$ Boltzmann's constant.  
It thus appears that the topology of the network has a non-negligible
influence on the entropy of the hydrogen distribution over the
available lattice sites.

\begin{figure}
\vspace{-1.0cm}
\includegraphics[width= 8cm]{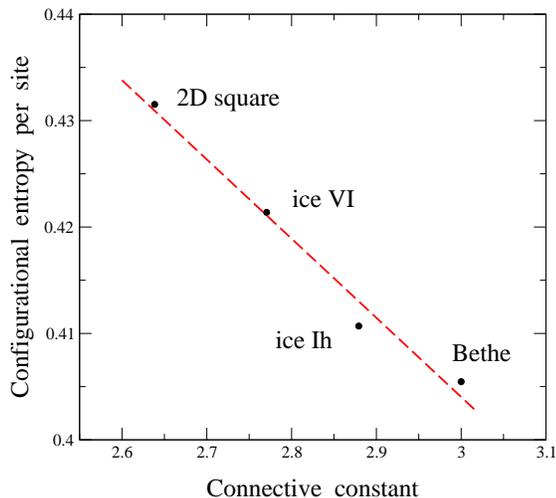}
\vspace{-0.3cm}
\caption{
Configurational entropy per site vs connective constant $\mu$ for
various networks. The dashed line is a least-square fit to the
data points.
}
\label{f6}
\end{figure}

In Fig.~6 we present the configurational entropy per site vs the
connective constant $\mu$ for ices Ih and VI, as well as for the
Bethe lattice (Pauling result) and the 2D square lattice.
The entropy value for the ice model on the square lattice has been 
taken from an exact calculation by Lieb \cite{li67}: $s$ = 0.4315.
One observes in this figure a negative correlation between the entropy
$s$ and $\mu$. The dashed line is a least-square fit to the data
points.
We do not see any reason why this correlation should be linear, but it
seems clear that there is a relation between both magnitudes. 
A qualitative argument for this relation is the following. 
The connective constant is a measure of the mean effective number of 
sites connected to a node in a long self-avoiding walk.  
For the distribution of hydrogen atoms on the links of the ice
networks according to the Bernal-Fowler rules, a larger $\mu$ means
an increase in the correlations in the occupancy of the available
hydrogen sites, as information on the actual occupancy of a link
propagates `more effectively' through the ice network.
Such an increase in the site-occupancy correlations causes a reduction
in the number of available configurations compatible with the 
ice rules on a given structure, which translates into a decrease in
the configurational entropy of the hydrogen distribution on the
considered ice network. 
In this line, relations between SAWs and order/disorder problems, 
such as the Ising model on lattices, have been found and studied in the 
past \cite{pe02,do70,he95b}.

Looking at Fig.~6 in more detail, one expects also that
dimensionality can play a role in the actual values of the
configurational entropy, and thus values derived for the 2D square
lattice and the Bethe lattice (without a well defined dimension $d$; 
for some purposes $d \to \infty$) may depart from the general trend 
found for 3D structures. This question should be investigated 
in the future, as configurational entropy associated to hydrogen
disorder can be precisely calculated for other crystalline ice
structures using methods as those described in 
Refs.~\cite{be07,he13}.

\section{Conclusions}

We have presented a calculation of self-avoiding walks on eight
crystalline ice structures, which include all topologically different
ice networks known so far.
SAWs have been enumerated up to a walk length $n = 27$, which 
allowed us to obtain the connective constants for the
different polymorphs. This parameter $\mu$ has turned out to be a
quantitative topological characteristic of ice structures. 

Correlations between the connective constant and other topological 
aspects of ice polymorphs have been found.
The constant $\mu$ is in fact related with the mean ring size
$\langle L \rangle$, although the
correlation between both quantities is largely influenced by the
minimum ring size $L_{min}$ in each structure.
A similar correlation between the connective constant and the 
parameter $a$ controlling the coordination sequences in these
crystalline structures ($N_k \sim a k^2$) has been found. This
network-dependent parameter gives the topological density of ice
polymorphs.

An interesting correlation has been found between the connective
constant and the configurational entropy of hydrogen-disordered  
ice polymorphs. A larger $\mu$ favors stronger correlations 
in hydrogen occupancy of crystal sites, thus causing a decrease in the
configurational entropy. This has been explicitly seen for ices Ih and
VI, and compared to other networks with coordination number $z = 4$,
such as the 2D square lattice and the Bethe lattice. 
A generalization to other crystalline ice polymorphs would be interesting,
to check the extent to which this correlation is followed 
for this kind of structures.
This will require a calculation of the configurational entropy for
other hydrogen-disordered ice polymorphs, which is expected
to change according to their topological characteristics.

\begin{acknowledgments}
R. Ram\'irez is thanked for inspiring discussions.
This work was supported by Direcci\'on General de Investigaci\'on (Spain) 
through Grant FIS2012-31713 
and by Comunidad Aut\'onoma de Madrid 
through Program MODELICO-CM/S2009ESP-1691.
\end{acknowledgments}

\end{document}